\documentclass[12pt]{article}
\usepackage{graphicx}
\usepackage{amssymb,amsmath,bm,bbm,paralist}

 \newcommand\pubnumber{FLAVOUR(267104)-ERC-86}
\newcommand\pubdate{\today}

\def\napoli{TUM Institute for Advanced Study\\
Lichtenbergstr. 2a, 85748 Garching, Germany}

\def\Title#1{\begin{center} {\Large #1 } \end{center}}
\def\Author#1{\begin{center}{ \sc #1} \end{center}}
\def\Address#1{\begin{center}{ \it #1} \end{center}}

\newcommand\pubblock{\rightline{\begin{tabular}{l} \pubnumber\\
         \pubdate  \end{tabular}}}
\newenvironment{Abstract}{\begin{quotation}  }{\end{quotation}}
\newenvironment{Presented}{\begin{quotation} \begin{center} 
             PRESENTED AT\end{center}\bigskip 
      \begin{center}\begin{large}}{\end{large}\end{center} \end{quotation}}
\def\Acknowledgements{\bigskip  \bigskip \begin{center} \begin{large}
             \bf ACKNOWLEDGEMENTS \end{large}\end{center}}

\def\beq{\begin{equation}}
\def\eeq#1{\label{#1}\end{equation}}
\def\eeqn{\end{equation}}

\def\beqa{\begin{eqnarray}}
\def\eeqa#1{\label{#1}\end{eqnarray}}
\def\eeqan{\end{eqnarray}}

\let\bar=\overbar

\def\Dslash{\not{\hbox{\kern-4pt $D$}}}
\def\dslash{\not{\hbox{\kern-2pt $\del$}}}

\def\msb{{\bar{\ssstyle M \kern -1pt S}}}

\begin{document}
\begin{titlepage}
\pubblock

\vfill
\Title{News on $\boldmath{B\to K^{(*)}\nu\bar\nu}$ in the Standard Model and beyond}
\vfill
\Author{ Jennifer Girrbach-Noe}
\Address{\napoli}
\vfill
\begin{Abstract}
An analysis of the rare $B$ decays $B\to K^{(*)}\nu\bar\nu$ is presented both
within the SM and beyond. The SM predictions for the branching ratios are updated and uncertainties reduced. For the NP analysis both a 
model independent approach is used and concrete
NP models are studied. The relations between $b\to s\nu\bar\nu$ and $b\to s\ell^+\ell^-$ transitions are analysed in detail. 
\end{Abstract}
\vfill
\begin{Presented}
Presented at the 8$^{\rm th}$ International Workshop on the CKM Unitarity Triangle (CKM 2014),\\
Vienna, Austria, September 8-12, 2014
\end{Presented}
\vfill
\end{titlepage}
\def\thefootnote{\fnsymbol{footnote}}
\setcounter{footnote}{0}

\section{Introduction}

The decays $B\to K^{(*)}\nu\bar\nu$ are theoretically very clean and will play a key role in the tests of the SM and
its extensions. They are especially sensitive to $Z$ penguins whereas $b\to s\ell^+\ell^-$ transitions are additionally
sensitive to dipole and scalar operators. Due to their sensitivity to right-handed couplings $B\to K^{(*)}\nu\bar\nu$ decays 
offer a powerful test of MFV. This talk is based on~\cite{Buras:2014fpa} and studies $B\to K^{(*)}\nu\bar\nu$ both
within the SM and beyond. In 2009 these decays were analysed in \cite{Altmannshofer:2009ma} but since the flavour precision era
is ahead of us and due to new lattice calculations and new data on $b\to s\ell^+\ell^-$ transitions it is time to have
a closer look at $B\to K^{(*)}\nu\bar\nu$ again. The main novelties of~\cite{Buras:2014fpa} are:
\begin{itemize}
 \item The SM predictions for the branching ratios and the angular observable $F_L$ are updated. Due to new lattice calculations
 the form factor uncertainties decreased considerably \cite{Bouchard:2013eph,Horgan:2013hoa}.
 \item We exploit the SU(2)$_L$ symmetry in order to relate $b\to s\ell^+\ell^-$ and $b\to s\nu\bar\nu$ transitions. 
\item New data on $B\to K^{*}\mu^+\mu^-$  and its impact on $B\to K^{(*)}\nu\bar\nu$ are included. 
\item We study correlations both in a model-independent approach and in concrete models beyond the SM.
\item The impact of lepton flavour non-universality is also discussed in~\cite{Buras:2014fpa} but not covered in this talk. 
\end{itemize}

\section{SM results}

In the SM only one operator contributes to $B\to K^{(*)}\nu\bar\nu$:
\begin{subequations}
\begin{align} 
{\mathcal{H}}_{\text{eff}}^\text{SM} &= - \frac{4\,G_F}{\sqrt{2}}V_{tb}V_{ts}^*
C_L^\text{SM} \mathcal O_L
~+~ \text{h.c.} \,,\quad \mathcal{O}_{L} =\frac{e^2}{16\pi^2}
(\bar{s}  \gamma_{\mu} P_L b)(  \bar{\nu} \gamma^\mu(1- \gamma_5) \nu)\\
 C_L^\text{SM}& = -X_t/s_w^2
\,,\quad 
X_t=1.469\pm0.017
\end{align}
\end{subequations}
and its Wilson  coefficient $C_L^\text{SM}$ is known very precisely (including NLO QCD corrections 
\cite{Buchalla:1993bv,Misiak:1999yg,Buchalla:1998ba} and two-loop electroweak contributions \cite{Brod:2010hi}).
The form factors for the exclusive decays $B\to K^{(*)}\nu\bar\nu$ have to be calculated by means of non-perturbative methods.
Great progress has been made by lattice calculation, especially at large $q^2$ \cite{Bouchard:2013eph,Horgan:2013hoa}. At
low $q^2$ the results from light-cone sum rules are used \cite{Ball:2004ye,Ball:2004rg}. Further details can be found
in~\cite{Buras:2014fpa} and the rescaled form factors are shown there in appendix~A. Our new result for the total branching ratios in the 
SM and for the $K^*$ longitudinal polarization fraction $F_L$ that update earlier determinations 
in~\cite{Altmannshofer:2009ma,Bartsch:2009qp} are given as:
\begin{align}
\text{BR}(B^+\to K^+\nu\bar\nu)_\text{SM} & =  (4.20 \pm 0.33 \pm 0.15) \times 10^{-6}, \\
\text{BR}(B^0\to K^{* 0}\nu\bar\nu)_\text{SM} & =  (9.93\pm0.74\pm0.35) \times 10^{-6}, \\
F^\text{SM}_L &= 0.53 \pm 0.05\,,
\end{align}
where the first error is due to form factor uncertainties and the second one  is
parametric and dominated by CKM uncertainties.
Our result for $\text{BR}(B^0\to K^{* 0}\nu\bar\nu)_\text{SM}$ is
roughly 40\% higher than the one presented in~\cite{Altmannshofer:2009ma} (for further details see~\cite{Buras:2014fpa}).
These numbers should be compared with the current experimental upper bounds.
BaBar finds the following upper bound~\cite{Lees:2013kla}:
\begin{align}
\text{BR}(B^+ \to K^+\nu\bar\nu) &<1.7\times 10^{-5}\ \text{(90\%\, CL)}.
\end{align}
Belle has the strongest bound on the $B\to K^*\nu\bar\nu$
\cite{Lutz:2013ftz}:
\begin{align}
\text{BR}(B^0 \to K^{*0}\nu\bar\nu) &<5.5\times 10^{-5}\ \text{(90\%\, CL)},
\\
\text{BR}(B^+ \to K^{*+}\nu\bar\nu) &<4.0\times 10^{-5}\ \text{(90\%\, CL)}.
\end{align}
Thus, currently the experimental bounds are roughly a factor of four larger than the SM predictions.

\section{Going beyond the SM}

\subsection{Effective field theory approach}
If we go beyond the SM there is one additional operator at low energy, namely $\mathcal{O}_{R}^\ell =\frac{e^2}{16\pi^2}(\bar{s}  
\gamma_{\mu} P_R b)(  \bar{\nu}_\ell \gamma^\mu(1- \gamma_5) \nu_\ell)$ with $\ell = e,\mu,\tau$. To be more general  we distinguish here between the 
different lepton flavours. Denoting $C_R^{\ell}$ the corresponding Wilson coefficient and analogously $C_L^{\ell}$ we define
  \begin{equation} 
 \epsilon_\ell = \frac{\sqrt{ |C_L^\ell|^2 + |C_R^\ell|^2}}{|C_L^\text{SM}|}~ \qquad
 \eta_\ell = \frac{-\text{Re}\left(C_L^\ell C_R^{\ell *}\right)}{|C_L^\ell|^2 + |C_R^\ell|^2}\,
\end{equation}
with $\epsilon_\ell = 1$ and $\eta_\ell = 0$ in the SM.
The changes with respect to the SM results depend only on these two quantities and can be expressed as follows:
\begin{align}
 \mathcal{R}_K \equiv \frac{\mathcal{B}_K}{\mathcal{B}_K^\text{SM}}   & =  \frac{1}{3}\sum_\ell (1 - 
2\,\eta_\ell)\epsilon_\ell^2\quad\xrightarrow{\text{LFU}} \quad(1 - 2\,\eta)\epsilon^2
 \,, \\
 \mathcal{R}_{K^*} \equiv \frac{\mathcal{B}_{K^*}}{\mathcal{B}_{K^*}^\text{SM}} 
  & =\frac{1}{3}\sum_\ell (1 +  \kappa_\eta \eta_\ell)\epsilon_\ell^2\quad\xrightarrow{\text{LFU}}\quad
(1 +  \kappa_\eta \eta)\epsilon^2
  \,, \\
 \mathcal{R}_{F_L} \equiv \frac{F_L}{F_L^\text{SM}} 
 & = \frac{\sum_\ell \epsilon_\ell^2(1 + 2 \,\eta_\ell)}{\sum_\ell \epsilon_\ell^2(1 + \kappa_\eta 
\eta_\ell)}\quad\xrightarrow{\text{LFU}}\quad\frac{1+2\eta}{1+\kappa_\eta\eta}
 \,.
\end{align}
Here we also show what happens in case of lepton flavour universality (LFU). As one can see these observables are
very sensitive to right-handed currents and are thus a powerful test of MFV.

In order to exploit the SU(2)$_L$ symmetry that connects left-handed charged leptons and neutrinos  we list here all relevant
dimension-6 operators that are invariant under the SM gauge group G$_\text{SM}$ that contribute to both  $b\to s\ell^+\ell^-$ and $b\to 
s\nu\bar\nu$ transitions \cite{Buchmuller:1985jz,Grzadkowski:2010es}\footnote{Such an EFT approach has recently received increasing 
interest in  flavour physics, see e.g.~\cite{Crivellin:2014zpa,Alonso:2014csa,Hiller:2014yaa}.}: 
\begin{subequations}
  \begin{align}
 Q_{Hq}^{(1)} &= i (\bar q_L \gamma_\mu q_L) H^\dagger D^\mu H
\,,&
 Q_{q\ell}^{(1)} &= (\bar q_L \gamma_\mu q_L) (\bar \ell_L\gamma^\mu \ell_L)
\,,\\
 Q_{Hq}^{(3)} &= i (\bar q_L \gamma_\mu\tau^a q_L) H^\dagger D^\mu\tau_a H
\,,&
 Q_{q\ell}^{(3)} &= (\bar q_L \gamma_\mu \tau^a q_L) (\bar \ell_L\gamma^\mu \tau_a\ell_L)
\,,\\
 Q_{Hd} &= i (\bar d_R \gamma_\mu d_R) H^\dagger D^\mu H
\,,&
 Q_{d\ell} &= (\bar d_R \gamma_\mu d_R) (\bar \ell_L\gamma^\mu \ell_L)
\end{align}
\end{subequations}
Furthermore we need to consider those that contribute only to $b\to s\ell^+\ell^-$ but not to $b\to s\nu\bar\nu$:
\begin{align}
 Q_{de} &= (\bar d_R \gamma_\mu d_R) (\bar e_R\gamma^\mu e_R)
\,,&
 Q_{qe} &= (\bar q_L \gamma_\mu q_L) (\bar e_R\gamma^\mu e_R)
\,.
\label{eq:ops2}
\end{align}
The corresponding Wilson coefficients are denoted as $c_{q\ell}^{(1)}$ etc. 
We do not include dipole and scalar operators here for simplicity. Dipole operators are only relevant at low $q^2$ and scalar operators
only in $B_s\to\mu^+\mu^-$.

\begin{table}[t]
\centering
\renewcommand{\arraystretch}{1.3}
\begin{tabular}{c|ccccccc}
\hline
model &
$\widetilde{c}_Z$ &
$\widetilde{c}_{q\ell}^{(1)}$ &
$\widetilde{c}_{q\ell}^{(3)}$ &
$\widetilde{c}_{qe}$ &
$\widetilde{c}_Z'$ &
$\widetilde{c}_{de}$ &
$\widetilde{c}_{d\ell}$
\\
\hline
MFV, $U(2)^3$ & x & x & x & x & - & - & -\\
MSSM & x & -& - & - & - & - & - \\
Part. comp.: bidoublet & x & - & - & - & - & - & - \\
Part. comp.: triplet & - & - & - & - & x & - & - \\
331 model  & x & x& - & x & - & - & -\\
general $Z^\prime$  model   & - & x & - & x & - & x & x\\
\hline 
\end{tabular}
\caption{Wilson coefficients that are sizable (x) and negligible/zero (-) in different NP models.}
\label{tab:nptable}
\end{table}

After electroweak symmetry breaking these Wilson coefficients can be mapped onto the basis of the usual $\Delta F=1$ 
operators. With a proper choice of normalization  we get (for details see~\cite{Buras:2014fpa}):
\begin{align}
{B\to K^{(*)}\nu\bar\nu}:\quad & C_L = C_L^\text{SM} +  \widetilde{c}_{q\ell}^{(1)}- \widetilde{c}_{q\ell}^{(3)} +  
\widetilde{c}_Z
\,,&
C_R &= \widetilde{c}_{d\ell} + \widetilde{c}_Z'
\label{eq:WC1}
\,,\\
{B\to K^{(*)}\ell^+\ell^-}:\quad & C_9 =C_9^\text{SM} + 
{\widetilde{c}_{qe}}+\widetilde{c}_{q\ell}^{(1)}+\widetilde{c}_{q\ell}^{(3)} -\zeta \,\widetilde{c}_Z
\,,&
C_9' &={\widetilde{c}_{de}} +\widetilde{c}_{d\ell} -\zeta \,\widetilde{c}_Z'
\,,\\
{B_s\to \mu^+\mu^-}:\quad&C_{10} =C_{10}^\text{SM} +{\widetilde{c}_{qe}}-\widetilde{c}_{q\ell}^{(1)} - 
\widetilde{c}_{q\ell}^{(3)} + \widetilde{c}_Z
\,,&
C_{10}' &= {\widetilde{c}_{de}} -\widetilde{c}_{d\ell}  + \widetilde{c}_Z'
\end{align}
In complete generality,  NP effects  in $b\to s\nu\bar\nu$ cannot be constrained by $b\to s \ell^+\ell^-$. However in most models
not all operators appear and thus only a subset of Wilson coefficients are non-zero. In tab.~\ref{tab:nptable} we summarize
which Wilson coefficients can appear in different NP models.

\begin{figure}[t]
\centering
 \includegraphics[width = 0.45\textwidth]{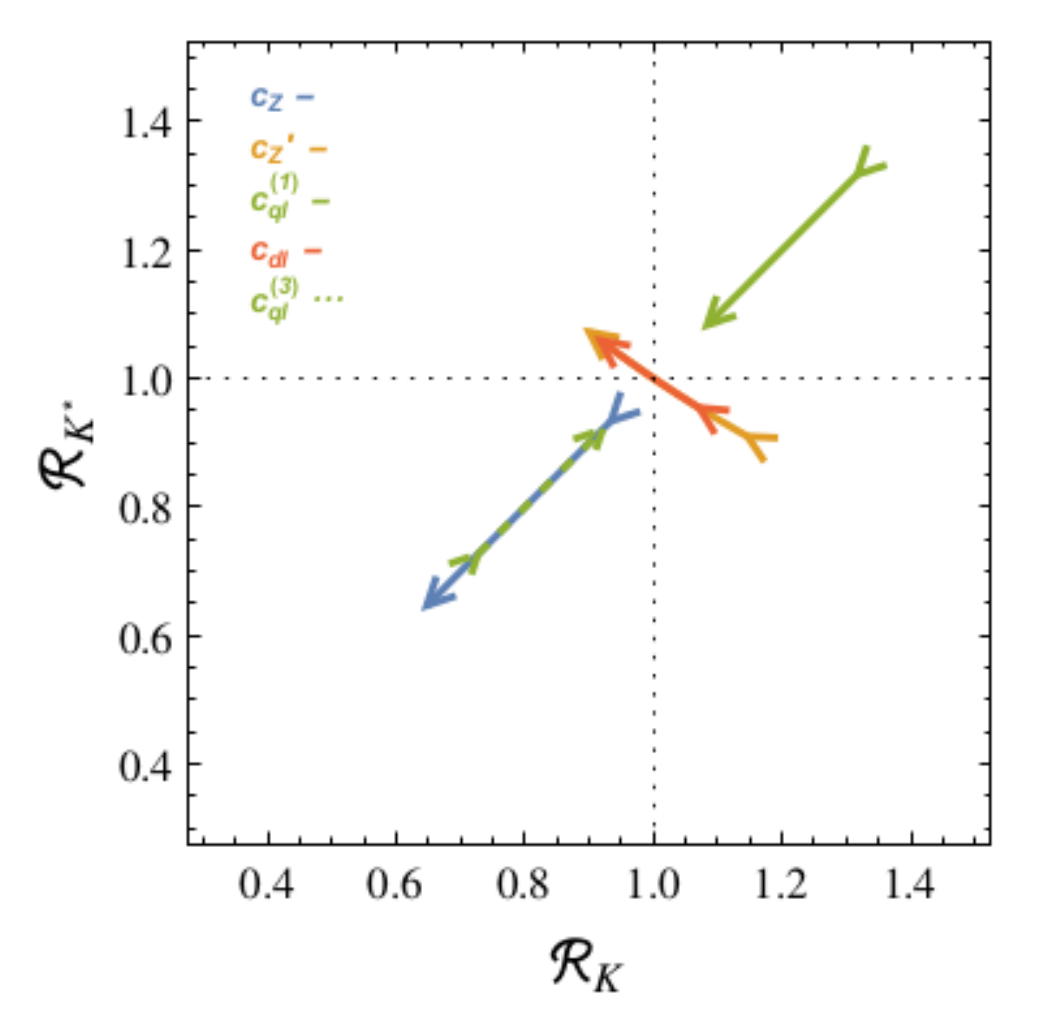}
\includegraphics[width=0.43\textwidth]{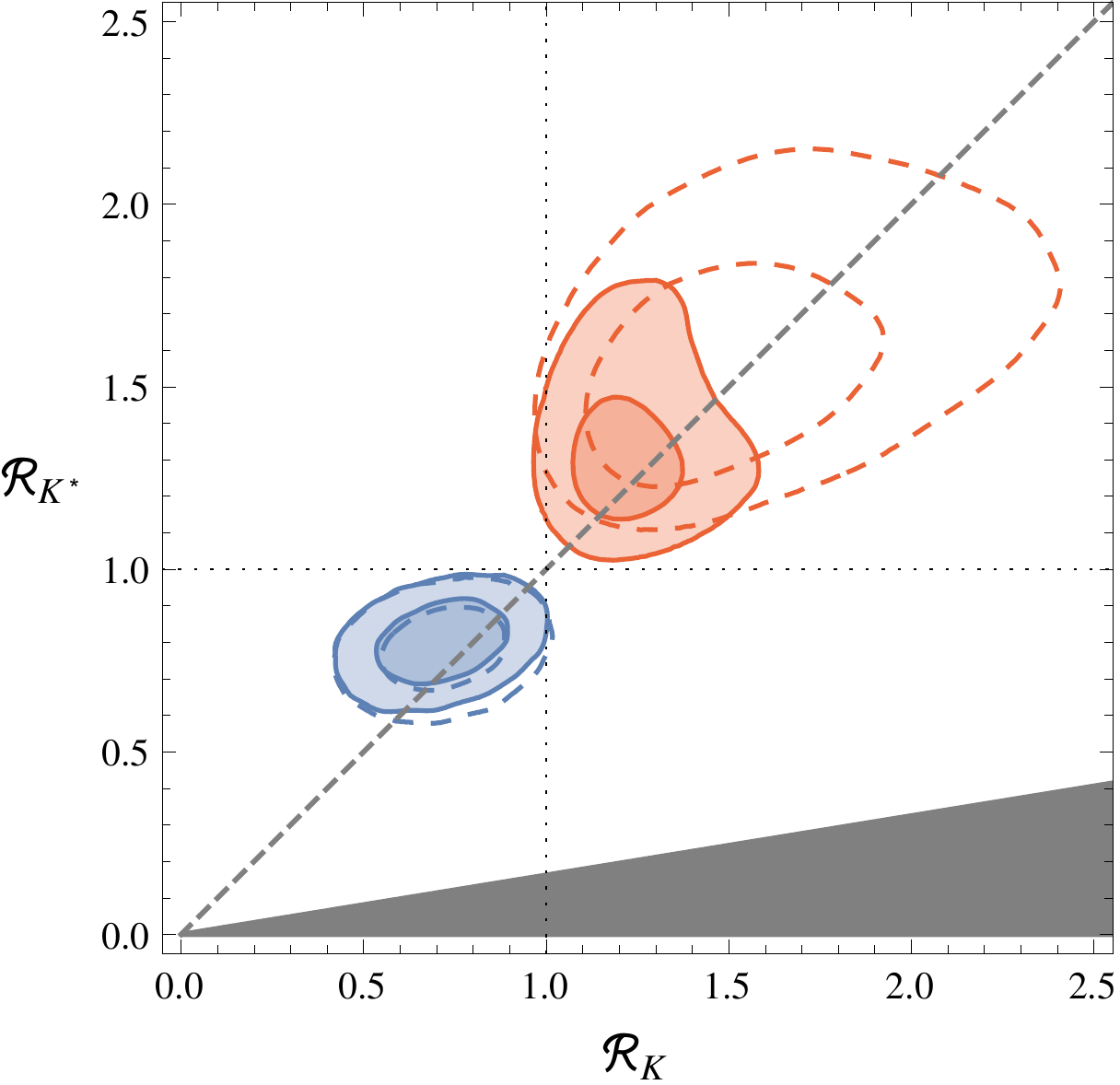}
\caption{Correlation between $\mathcal{R}^*$ and $\mathcal{R}$. Left: only one Wilson coefficient is non-zero (blue: {$\tilde c_Z$}, 
yellow: {$\tilde c_Z'$}, green: {$\tilde c_{ql}^{(1)}$} (solid) or
{$\tilde c_{ql}^{(3)}$} (dashed), red: {$\tilde c_{dl}$}). Right: only $Z$ penguins, i.e. $\widetilde{c}_Z$ and 
$\widetilde{c}_Z^\prime$ (blue) and  only 4-fermion operators, i.e. $c_{ql}^{(1)}$, $c_{qe}$, $c_{d\ell}$, $c_{de}$ (red) with  real 
(solid) 
or complex (dashed) Wilson coefficients, respectively.}
\label{fig:RKvsRKstar}
\end{figure}

In our numerical analysis we include bounds from $b\to s\ell^+\ell^-$ transitions that mainly constrain $C_{9,10}^{(')}$  using 
a global numerical analysis from
\cite{Altmannshofer:2011gn,Altmannshofer:2013foa,AS2014,Wolfitalk}\footnote{Similar global analyses have been performed 
in \cite{Descotes-Genon:2013wba,Beaujean:2013soa}. An analysis of $B\to K^{*}\ell^+\ell^-$ in Randall Sundrum was performed in~\cite{Biancofiore:2014wpa}.}. Since this is based on data from $b\to 
s\mu^+\mu^-$ transitions the bounds change in case of lepton flavour non-universality. 
On the left-hand side of fig.~\ref{fig:RKvsRKstar} we show the correlation between $\mathcal{R}^*$ and $\mathcal{R}$ where only one
Wilson coefficient is non-zero. The coloured arrows correspond to the $2\sigma$ allowed ranges of the $b\to 
s\mu^+\mu^-$ data, with the direction of the arrow 
pointing from negative to positive values for the $\tilde c_i$. On the right-hand side we show the result where we assume NP only
in $Z$ penguins, thus $\widetilde{c}_Z$ and 
$\widetilde{c}_Z^\prime$ are non-zero (blue) and where NP only appears in 4-fermion operators (red), corresponding to a $Z^\prime$ model.
It is interesting to note that $B\to K^{(*)}\nu\bar\nu$ are both enhanced
   for $Z^\prime$ and both suppressed for $Z$ which makes it possible to distinguish between those two scenarios.

\subsection{Concrete NP models}

Here we sketch our results in concrete NP models: $Z^\prime$ models,  331 model, MSSM and Partial Compositeness. Leptoquark models
and the case of lepton flavour non-universality are not covered in this talk but can be found in~\cite{Buras:2014fpa}\footnote{A recent study of
$b\to s\nu\bar\nu$ transitions in RS$_c$ models can be found in~\cite{Biancofiore:2014uba}.}.

\subsubsection{General $Z^\prime$ models}
\begin{figure}[ptb]
\centering
\includegraphics[height=0.9\textheight]{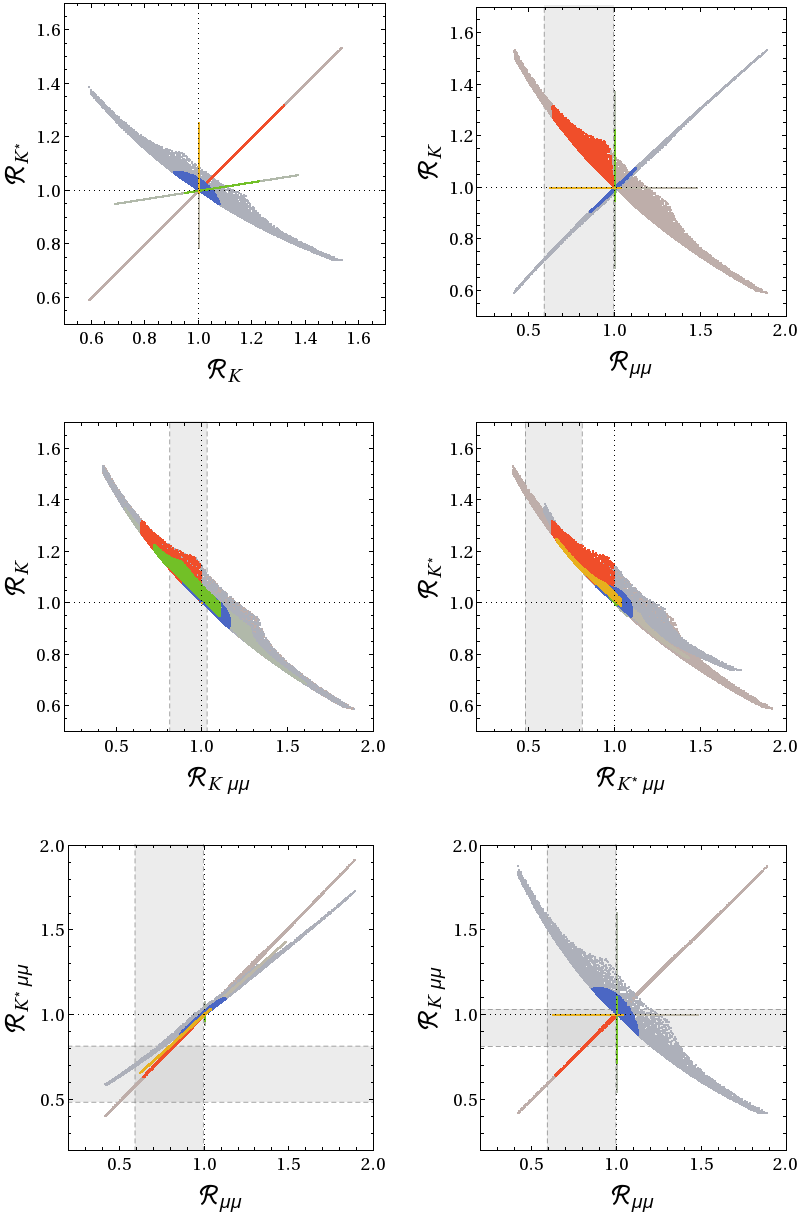}
\caption{Various correlations between observables in  LHS (red), RHS (blue), LRS (green), ALRS (yellow), assuming LFU and 
$\Delta_R^{\nu\nu} 
= \Delta_R^{\ell\ell}=0$. All points satisfy  $0.9\leq C_{B_s}\leq 1.1$, $-0.14\leq S_{\psi\phi}\leq 0.14$.
Grey  regions are disfavoured at $2\sigma$ by $b\to s\mu^+\mu^-$ constraints.}
\label{fig:Zprime}
\end{figure}

We will first study general $Z^\prime$ models as discussed in detail in~\cite{Buras:2012jb,Buras:2013qja,Buras:2014zga}.
We assume that NP contributions 
are dominated by the tree-level exchange of a $Z^\prime$ that transforms as a singlet under $SU(2)_L$. We distinguish here between
four different scenarios in 
which only LH quark couplings are present (LHS, red), the one with only RH couplings 
(RHS, blue), the one with LH and RH couplings being equal (LRS, green) and one with 
these couplings differing by sign (ALR, yellow). In fig.~\ref{fig:Zprime} we show the result for several correlations defining
\begin{align}
 \mathcal R_{\mu\mu} &= \frac{\text{BR}(B_s\to\mu^+\mu^-)}{\text{BR}(B_s\to\mu^+\mu^-)_\text{SM}}
 \,,\\
 \mathcal R_{K\mu\mu} &= \frac{\text{BR}(B^+\to K^+\mu^+\mu^-)^{[15,22]}}{\text{BR}(B^+\to K^+\mu^+\mu^-)^{[15,22]}_\text{SM}}
 \,,\quad
 \mathcal R_{K^*\mu\mu}= \frac{\text{BR}(B^0\to K^{*0}\mu^+\mu^-)^{[15,19]}}{\text{BR}(B^0\to K^{*0}\mu^+\mu^-)^{[15,19]}_\text{SM}}
 \,,
\end{align}
where 
constraints
from $\Delta F = 2$ observables and $b\to s\mu^+\mu^-$ constraints are included.  Interestingly, the present suppressions in the data in 
$B_s\to\mu^+\mu^-$, $B\to K^{(*)} \mu^+\mu^-$   favour left-handed currents that can be explained by $Z$ (tree or penguins) and 
$Z^\prime$. As shown in fig.~\ref{fig:RKvsRKstar} $B\to K^{(*)}\nu\bar\nu$ can distinguish these two mechanisms. 
Using the DNA charts in~\cite{Buras:2013ooa} a qualitative understanding of these plots can be gained.

\subsubsection{331 models}

\begin{figure}
\centering
\includegraphics[width=0.45\textwidth]{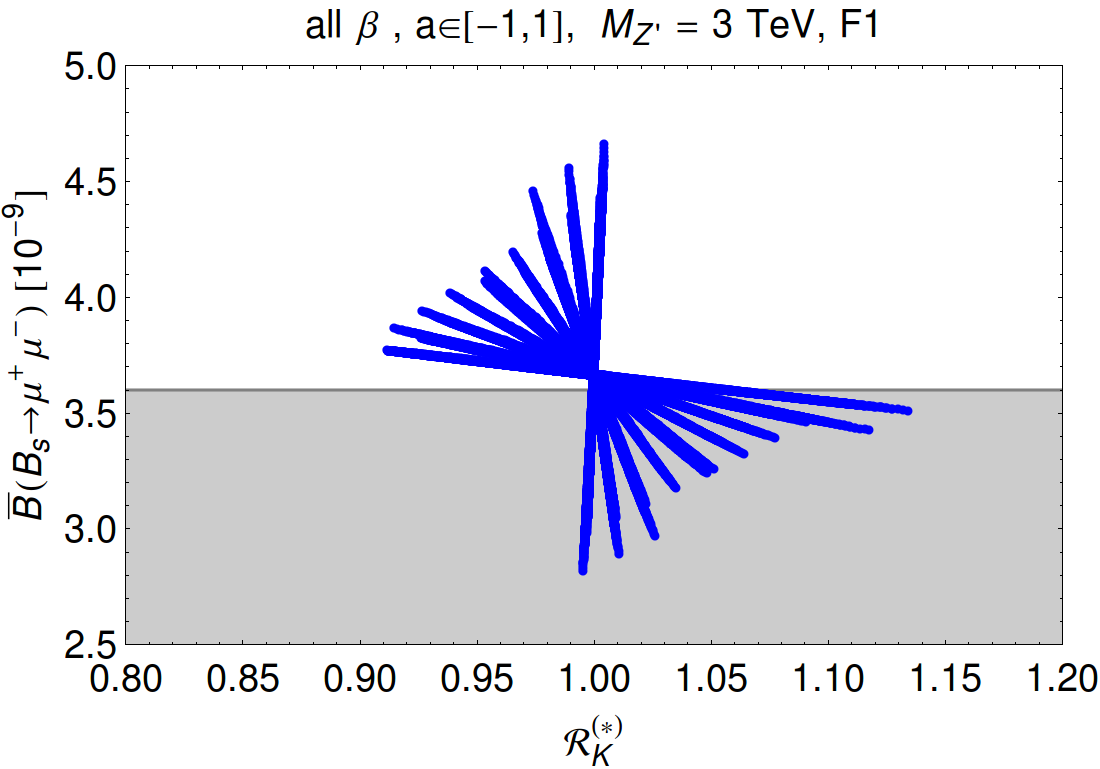}
\includegraphics[width=0.45\textwidth]{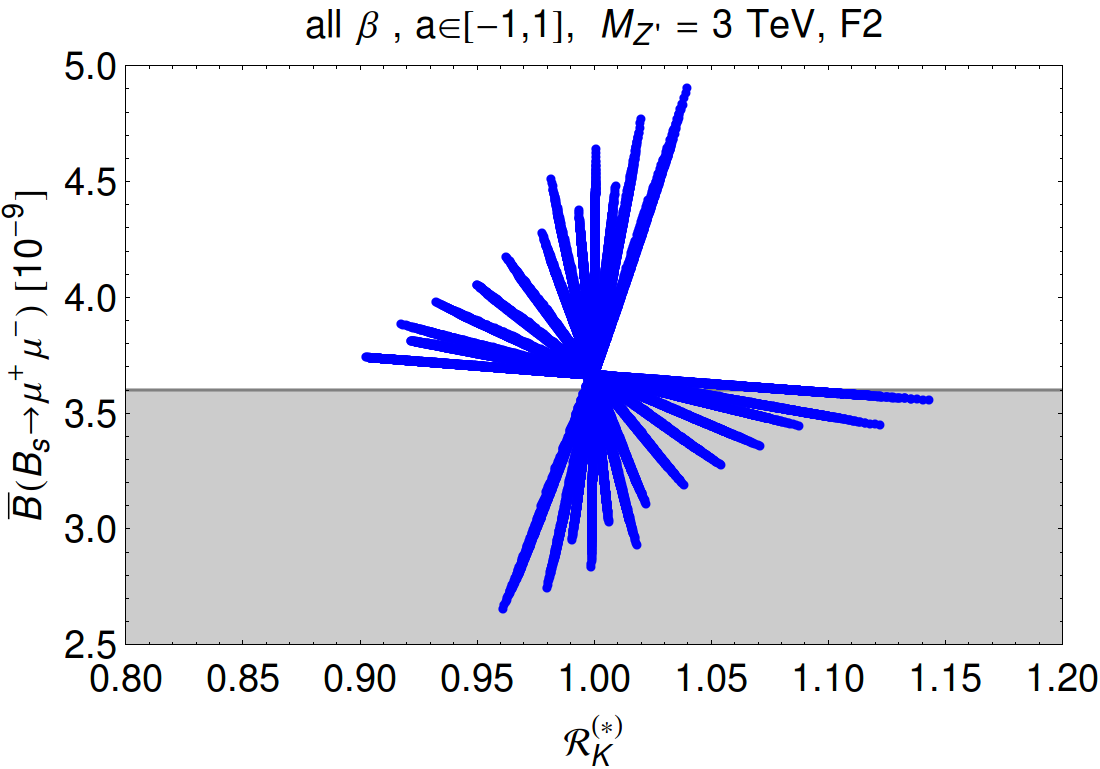}
\caption{Correlation
 $\overline{\text{BR}}(B_s\to\mu^+\mu^-)$ versus $\mathcal{R}_K^{(*)}$ in the 331 model for $M_{Z^\prime} = 3~$TeV for fermion 
representations F1 (left) and F2 (right).}
\label{fig:BsmuvsBK}
\end{figure}

The 331 model is based on the symmetry group $SU(3)_C\times SU(3)_L\times U(1)_X$ and is a concrete realization of a $Z^\prime$ model. A 
nice theoretical feature of this model is that there is an explanation of why there are exactly three generations. This follows from 
anomaly cancellation and asymptotic freedom of QCD. In the 331 model the $Z^\prime$ has flavour changing left-handed  quark currents at 
tree level while right-handed currents are flavour conserving. Consequently we have $\mathcal{R}^\star=\mathcal{R}$. Due to $Z^\prime-Z$ 
mixing also $Z$ flavour changing couplings at tree level are induced. Details on the model can be found 
in~\cite{Buras:2014yna,Buras:2013dea,Buras:2012dp}. The relevant Wilson coefficients are    $\widetilde{c}_Z$, 
$\widetilde{c}_{ql}^{(1)}$ and $\widetilde{c}_{qe}$  with $\widetilde{c}_{q\ell}^{(1)}\propto\widetilde{c}_Z$. There are several different
versions of 331 models depending on the parameter $\beta$ that determines the charge operator. In fig.~\ref{fig:BsmuvsBK} we scanned over
many different versions and show the possible range in the $\overline{\text{BR}}(B_s\to\mu^+\mu^-)-\mathcal{R}_K$ region for the two 
different fermion representations that are possible in 331 models. Constraints from $\Delta F = 2$
observables ($\Delta M_s$ and $S_{\psi\phi}$) and from $b\to s\mu^+\mu^-$ transitions are included as well as bounds from 
electroweak precision observables.  Since
$\widetilde{c}_{qe}+\widetilde{c}_{q\ell}^{(1)}$ enters $C_9$ while  $\widetilde{c}_{qe}-\widetilde{c}_{q\ell}^{(1)}$ enters 
$C_{10}$ it is difficult to get large effects in $B_d\to K^*\mu^+\mu^-$ and $B_s\to\mu^+\mu^-$ simultaneously. A suppression of 
$B_s\to\mu^+\mu^-$, as favoured by present data, 
almost always implies an enhancement of $\mathcal{R}_K$, but at most by 15\%. Models where both are enhanced or suppressed
simultaneously are excluded due to constraints from electroweak observables.

\subsubsection{MSSM}

\begin{figure}
\centering
\includegraphics[width=0.4\textwidth]{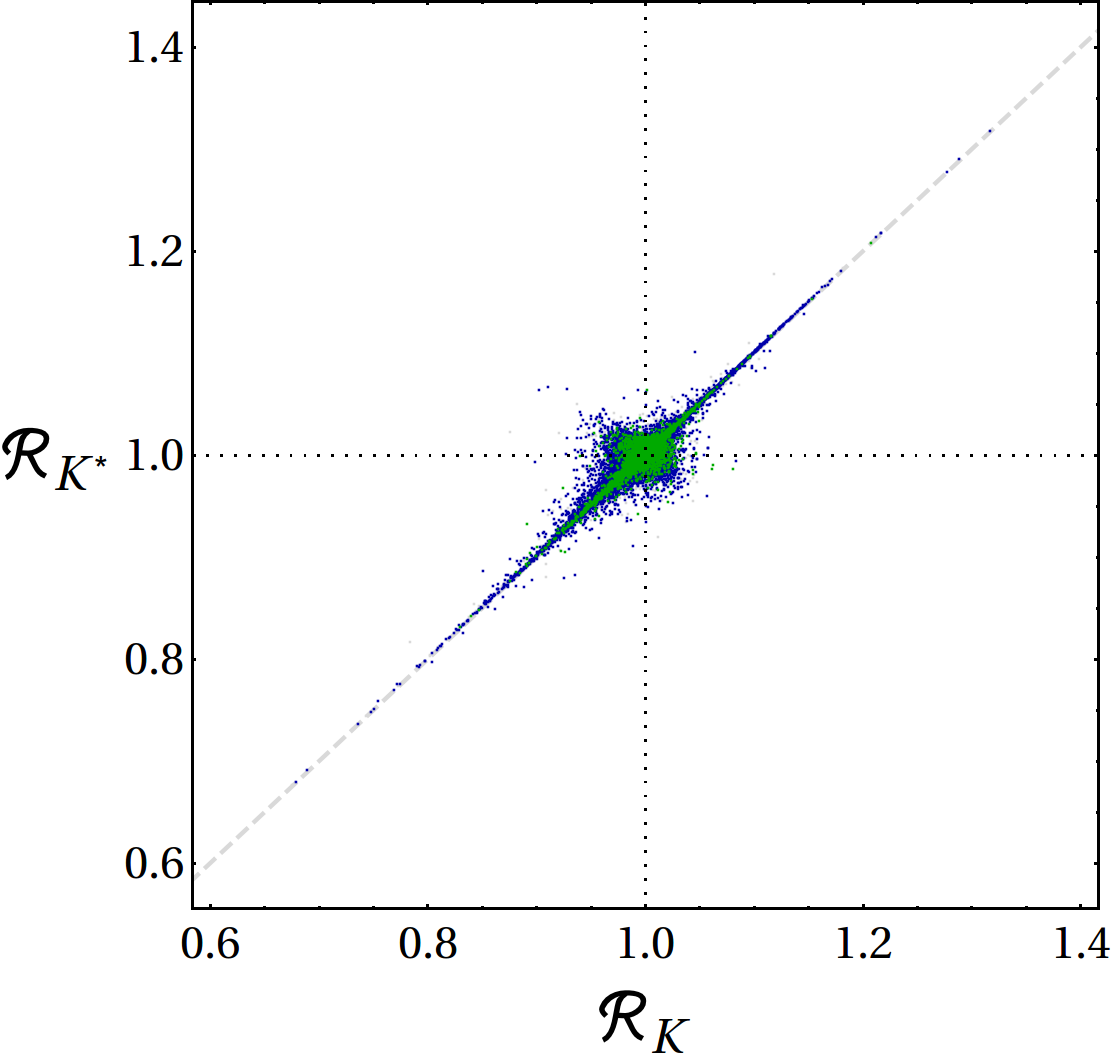}
\includegraphics[width=0.4\textwidth]{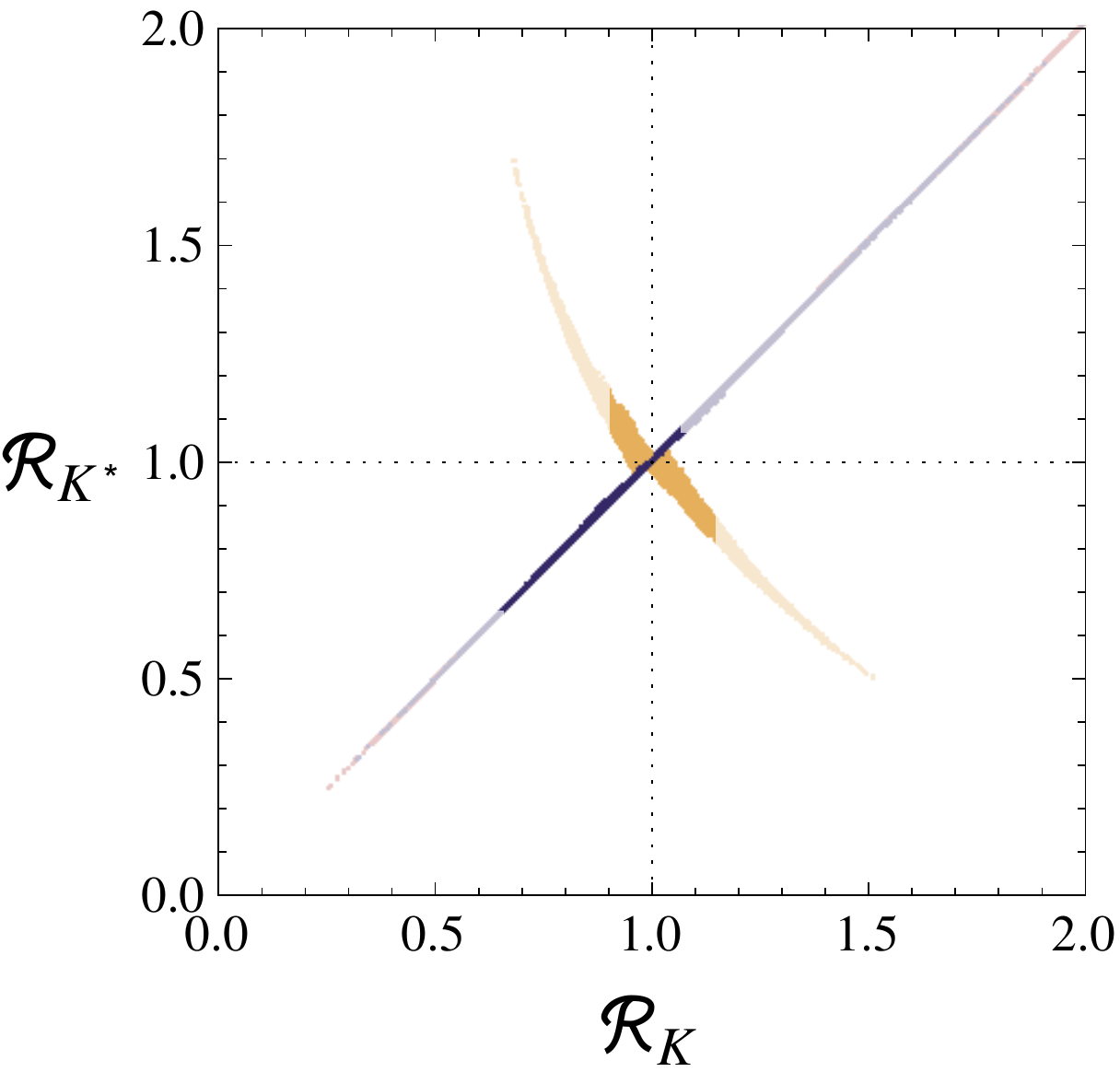}
\caption{Left: Allowed ranges for $\mathcal{R}^\star$ and $\mathcal{R}$ in the MSSM. All dark points pass flavour and collider constraints; 
black points have the corrected lightest Higgs mass. Right: Allowed ranges  in models with partial compositeness, for two different choices 
of fermion representations: bidoublet model (blue),  triplet model (yellow). Light 
points are disfavoured at $2\sigma$ by $b\to s\mu^+\mu^-$ data. Plot adapted from~\cite{Straub:2013zca}.}
\label{fig:RK-SUSY}
\end{figure}

In the MSSM the dominant effects arise from $Z$ penguins, thus $\widetilde{c}_Z$ (only large in non-MFV) and $\widetilde{c}_Z'$ (small due 
to constraints from $\text{BR}(B_s\to\mu^+\mu^-)$). The result of our parameter scan is shown in fig.~\ref{fig:RK-SUSY} (left). In our 
scan LHC bounds on sparticle masses are included using \texttt{FastLim 1.0}~\cite{Papucci:2014rja}, the lightest  neutralino is the LSP, 
\texttt{SUSY\char`_FLAVOR}~\cite{Crivellin:2012jv} is used to impose FCNC constraints and  the lightest Higgs mass  is computed with 
\texttt{SPheno 3.3.2} \cite{Porod:2003um,Porod:2011nf}. As one can see RH currents are small in the MSSM,
         so that $\mathcal{R}_K\approx\mathcal{R}_{K^*}$ and  $B\to K^{(*)}\nu\bar\nu$ are at most 30\% enhanced or suppressed.

\subsubsection{Partial Compositeness}

A simple model with partial compositeness was analysed in~\cite{Straub:2013zca} where the dominant contribution to $b\to s\nu\bar\nu$ 
transitions come
from tree-level flavour-changing $Z$ couplings. In the ``bidoublet model'' $\widetilde{c}_Z$ has the largest effect and $\widetilde{c}_Z'$ in the ``triplet model''.
The results of these two models are shown on the right hand side of fig.~\ref{fig:RK-SUSY}.

\section{Summary}

We analysed $b\to s\nu\bar\nu$ transitions within the SM and beyond. Due to reduced form factor uncertainties our updated SM predictions
have smaller errors of about 10\% and are also more reliable.  $\text{BR}(B\to K^*\nu\bar\nu)_\text{SM}$ is found to be by $40\%$ 
larger than previous estimates. Using the SU(2)$_L$ symmetry we correlate $b\to s\nu\bar\nu$ and $b\to s\ell^+\ell^-$ and impose
the constraints by recent measurements of $b\to s\mu^+\mu^-$ transition on $B\to K^{(*)} \nu\bar\nu$. This limits the relative deviations 
from the SM to roughly  $\pm60\%$ and in case of lepton flavour non-universality if NP only affects muons, the effects are at 
most $\pm20\%$. We find that  $b\to s\nu\bar\nu$ gives complementary information to NP in $b\to s\ell^+\ell^-$ and can help to disentangle
possible NP dynamics behind the anomalies in $B\to K^{(*)} \mu^+\mu^-$.

\Acknowledgements
I thank the organisers for the opportunity to give this talk and
my collaborators   A.~J.~Buras, C.~Niehoff and  D.~Straub for an enjoyable collaboration.  I thank Andrzej Buras and David Straub for proofreading this 
manuscript. I acknowledge financial support by ERC Advanced Grant
project ``FLAVOUR'' (267104).

\end{document}